\documentclass[prl,aps,twocolumn,showpacs]{revtex4-1}
\usepackage{graphicx}
\usepackage{amsmath} 
\usepackage{bm}
\usepackage{braket}
\usepackage{psfrag}  
\usepackage{latexsym}  
\usepackage{amstext}
\usepackage{amsxtra} 
\usepackage{float}
\usepackage{upgreek}  

\usepackage{times} 

\begin{document}

\newcommand{\To}{T_c^0}
\newcommand{\kB}{k_{\rm B}}
\newcommand{\dT}{\Delta T_c}
\newcommand{\lo}{\lambda_0}
\newcommand{\cs}{$\clubsuit \; $}
\newcommand{\thold}{t_{\rm hold}}
\newcommand{\Nmf}{N_c^{\rm MF}} 
\newcommand{\LM}{L_3^{\rm M}}
\newcommand{\Tmf}{T_c^{\rm MF}}
\newcommand{\downstate}{\left\vert\downarrow\right\rangle}  
\newcommand{\upstate}{\left\vert\uparrow\right\rangle}
\newcommand{\Ndown}{N_{\downarrow}}
\newcommand{\Nup}{N_{\uparrow}}

\title{
Connecting Berezinskii-Kosterlitz-Thouless and BEC phase transitions \\ by tuning interactions in a trapped gas
}

\author{Richard J. Fletcher$^\star$$^\dag$, Martin Robert-de-Saint-Vincent$^\star$$^\ddag$, \\ Jay Man, Nir Navon, Robert P. Smith, Konrad G. H. Viebahn, and Zoran Hadzibabic}
\affiliation{Cavendish Laboratory, University of Cambridge, J.~J.~Thomson Avenue, Cambridge CB3~0HE, United Kingdom}

\begin{abstract}
We study the critical point for the emergence of coherence in a harmonically trapped two-dimensional Bose gas with tuneable interactions. Over a wide range of interaction strengths we find excellent agreement with the classical-field predictions for the critical point of the Berezinskii-Kosterlitz-Thouless (BKT) superfluid transition. This allows us to quantitatively show, without any free parameters, that the interaction-driven BKT transition smoothly converges onto the purely quantum-statistical Bose-Einstein condensation (BEC) transition in the limit of vanishing interactions. 
\end{abstract}

\date{\today}

\pacs{67.85.-d, 03.75.Hh, 05.30.-d}


\maketitle

Reducing the dimensionality of a physical system increases the importance of thermal fluctuations and can profoundly affect the type of order that the system can display at low temperatures~\cite{Peierls:surprises,Bogoliubov:1960,Hohenberg:1967,Mermin:1966}. In a uniform two-dimensional (2d) Bose gas, the true long-range-order (LRO) associated with Bose-Einstein condensation (BEC) is precluded at any non-zero temperature by the Mermin-Wagner theorem.
However, an interacting 2d Bose gas still undergoes the Berezinskii-Kosterlitz-Thouless (BKT) transition to a superfluid state at a critical phase-space density $D_\text{BKT}$~\cite{Berezinskii:1971,Kosterlitz:1973}. 
Quantitatively predicting $D_\text{BKT}$ requires an accurate description of the non-perturbative behaviour of interacting bosons in the fluctuation region near the critical point. Classical-field simulations~\cite{Prokofev:2001} model this behaviour by a turbulent matter-wave field~\cite{Berloff:2009},  and predict $D_\text{BKT}=\ln\left(380/\tilde{g}\right)$, where $\tilde{g}$ is a dimensionless measure of the interaction strength~\cite{Note1}. 
This result makes it manifest that the transition is interaction driven; 
the critical temperature,  $T_c \propto n/D_\text{BKT}$, where $n$ is the gas density, vanishes in the non-interacting limit, $\tilde{g}\rightarrow 0$, for any non-infinite $n$.
While for phase-space density $D> D_\text{BKT}$ true LRO is still absent, the first order correlation function, $g_1(r)$, decays only algebraically at large distance, in contrast to the exponential decay in the normal degenerate gas. Such extended coherence is sufficient for superfluidity~\cite{Holzmann:2007a}, and in practice offers a signature of the phase transition~\cite{Bramwell:1994,Holzmann:2008a,Bisset:2009b, Hadzibabic:2011}.

In contrast to the infinite uniform system, in a 2d harmonic trap, pertinent to most ultracold-atom experiments on BKT physics~\cite{Hadzibabic:2006,Schweikhard:2007,Kruger:2007,Clade:2009,Tung:2010,Rath:2010,Plisson:2011,Hung:2011,Yefsah:2011,Desbuquois:2012,Choi:2012,Choi:2013,Ha:2013,Desbuquois:2014,Ries:2014}, the modified density of states allows for a BEC transition to occur in the ideal gas ($\tilde{g} =0$)~\cite{Masut:1979,Bagnato:1991} (see also~\cite{Fischer:2002b}). In an isotropic trap of frequency $\omega_r$, it should occur at a critical atom number~\cite{Masut:1979,Bagnato:1991}
\begin{equation} 
N_\text{c}^0=\frac{\pi^2}{6}\left(\frac{\kB T}{\hbar \omega_r}\right)^2 .
\label{eq:Nc2D}
\end{equation}
This ideal-gas BEC transition is similar to the familiar condensation in 3d; it is purely quantum-statistical and follows from Einstein's standard argument of the saturation of the excited states. 
However, the importance of dimensionality emerges if interactions are introduced. In 3d, ideal-gas BEC occurs when in the trap centre $D  \approx 2.612$; weak interactions slightly shift the critical atom number~\cite{Dalfovo:1999,Gerbier:2004,Smith:2011}, but do not alter the BEC-like nature of the transition. In 2d,  while $N_c^0$ is finite, the phase-space density required in the trap centre for the excited states to saturate is infinite, just as in a uniform system~\cite{Hadzibabic:2011}. For any $\tilde{g}>0$ this is unattainable and the excited states can accommodate any number of particles~\cite{Note2}. The BEC transition is thus suppressed and one expects it to be replaced by the BKT transition with non-infinite $D_\text{BKT}$~\cite{Holzmann:2007a}.

These arguments suggest that the two conceptually very different phase transitions, the interaction-driven BKT and the saturation-driven BEC, are in fact continuously connected as $\tilde{g} \rightarrow 0$~\cite{Holzmann:2007a,Hadzibabic:2008}. The harmonic trapping potential offers the opportunity to experimentally observe this unification of BKT and BEC physics. While in an infinite uniform gas no transition occurs for $\tilde{g}=0$, in a harmonic trap a transition always occurs at a finite $N_c$ and always results in significantly extended coherence of the gas. The nature of this transition is quantitatively encoded in the value of $N_c$.
This picture is supported by the calculations of the critical atom number $N_\text{c}^\text{BKT}$~\cite{Holzmann:2007a,Holzmann:2010}, based on the classical-field simulations~\cite{Prokofev:2001,Prokofev:2002}, which suggest that $N_\text{c}^\text{BKT}$ smoothly connects to $N_\text{c}^0$ in the limit $\tilde{g}=0$.
In a finite-size system the change from the BKT to the BEC transition is a crossover that spans a nonzero range of $\tilde{g}$ values, but for realistic experimental parameters the width of this crossover region is very small, $\tilde{g} \lesssim 10^{-2}$~\cite{Clade:2009,NoteCrossover}.

Various signatures of a phase transition, including emergence of extended coherence~\cite{Hadzibabic:2006,Kruger:2007,Clade:2009,Tung:2010,Plisson:2011} and superfluidity~\cite{Desbuquois:2012}, have been observed in trapped 2d gases with different specific $\tilde{g}$ values. On the other hand, systematic studies with tuneable $\tilde{g}$ have focussed on in-situ measurements of the equation of state~\cite{Hung:2011,Ha:2013}, which do not directly reveal any striking signatures of the infinite-order BKT transition. 

In this Letter, we systematically study the critical point for the emergence of extended coherence in a harmonically trapped 2d Bose gas over a wide range of interaction strengths, $0.05<\tilde{g}<0.5$. We show, without any free parameters, that $N_c$ generally agrees very well with the beyond-mean-field calculation of $N_\text{c}^\text{BKT}$~\cite{Holzmann:2010}, and converges onto $N_\text{c}^0$ of Eq.~(\ref{eq:Nc2D}) as $\tilde{g}\rightarrow 0$. The critical chemical potential, $\mu_c$, which directly reveals uniform-system conditions for a phase transition to occur in the trap centre, also agrees with the BKT theory and converges onto the BEC value, $\mu_c=0$, for $\tilde{g}\rightarrow 0$. Our measurements also reiterate the importance of the suppression of density fluctuations in the normal state near the BKT critical point, previously observed in~\cite{Clade:2009,Tung:2010,Hung:2011,Plisson:2011,Yefsah:2011}.

The experiment was carried out using a $^{39}$K gas, in the apparatus described in~\cite{Campbell:2010}. For 2d trapping, the tight axial (vertical) confinement is provided by two repulsive ``blades'' of blue-detuned light, formed by passing a 532-nm gaussian beam through a 0-$\pi$ phase-plate~\cite{Smith:2005,Rath:2010}, while a red-detuned 1064-nm dipole trap provides the in-plane (horizontal) confinement. The radial and axial trapping frequencies are $(\omega_r,\omega_z)\approx 2\pi \times (38,4100)$~Hz. For all our measurements  $T\in\left[140~\text{nK},190~\text{nK}\right]$ and $\mu/\kB < 100$~nK, resulting in a small $\left(<30\%\right)$ occupation of the excited axial states. 
The interaction strength, $\tilde{g}=\sqrt{8\pi}a/\ell_z$~\cite{Hadzibabic:2011}, where $a$ is the s-wave scattering length and $\ell_z=\sqrt{\hbar/(m\omega_z)}$, is controlled via a Feshbach resonance centred at $402.5$~G~\cite{Roati:2007,Campbell:2010}.

To characterise long-range coherence of a gas we study its (in-plane) momentum distribution $n(k)$~\cite{Tung:2010}. 
A change in the functional form of $g_1(r)$ leads to a dramatic change in its values at distances much larger than the thermal wavelength $\lambda=h/\sqrt{2\pi m\kB T}$~\cite{Hadzibabic:2011}, 
and an increase of coherence over some large distance $L$ manifests itself in enhanced population of the low-momentum states $k\lesssim 2\pi/L$. 
Thus, unlike the in-trap density distribution, which varies very smoothly through the BKT critical point~\cite{Clade:2009,Tung:2010,Yefsah:2011,Hung:2011}, $n(k)$ can provide a dramatic signature of the phase transition~\cite{Tung:2010}.

\begin{figure}[tbp] 
\includegraphics[width=1\columnwidth]{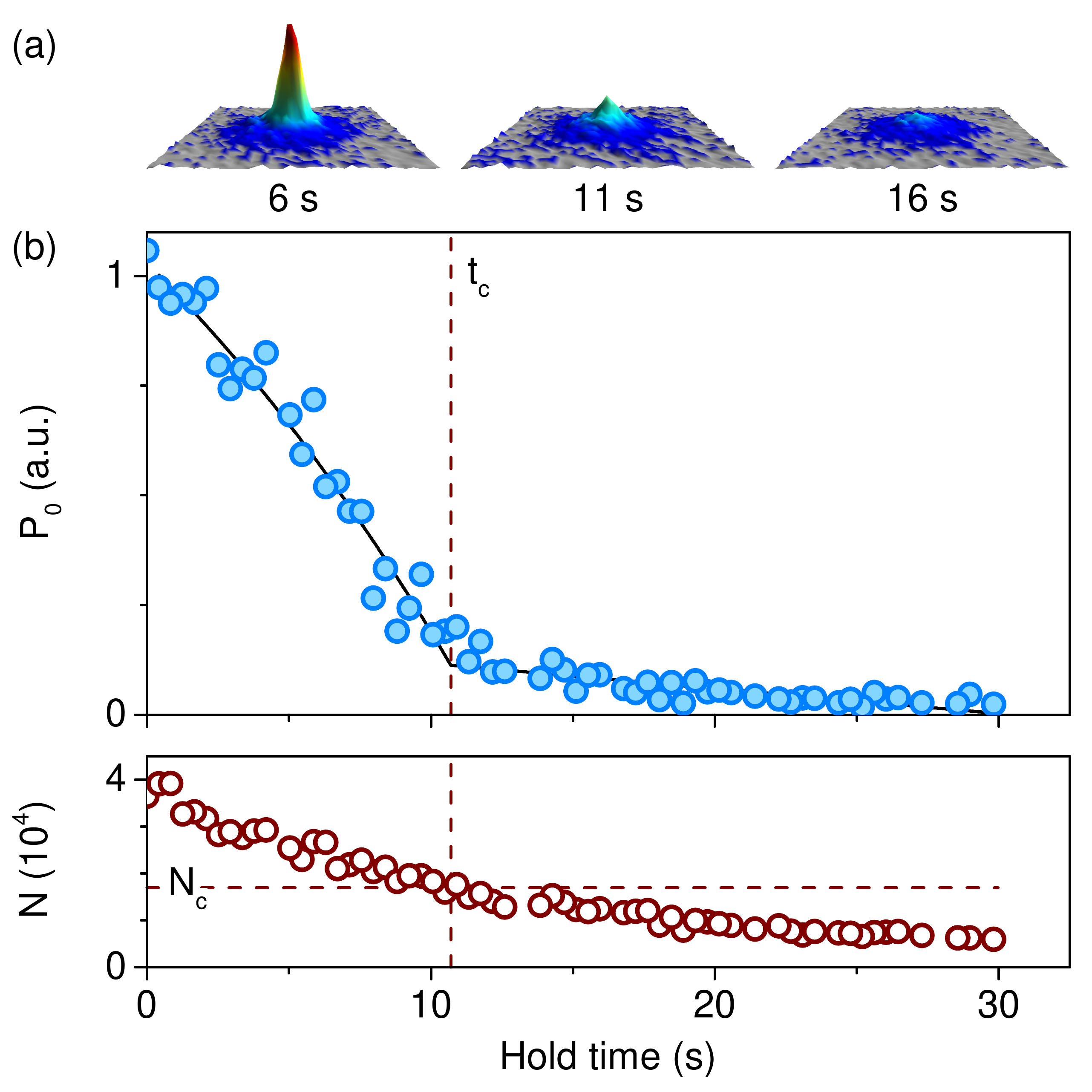}
\caption{(color online) Determination of the critical point for the onset of coherence, for  $\tilde{g}=0.28$ and $T\approx140$ nK. (a) Evolution of the momentum distribution, $n(k)$, with the hold time $t$ (see text). Extended coherence is revealed as a sharp peak in $n(k)$.   Each image is an average of three experimental realisations. (b) Evolution of the momentum-distribution peak, $P_0$, and the smoothly decaying total atom number $N$.  We associate the threshold-like behaviour of $P_0$ with the critical time $t_\text{c}$ and deduce the corresponding $N_\text{c}$. The solid line is a heuristic piecewise fit function used to determine $t_\text{c}$~\cite{Note4}.}
\label{fig:decay}
\end{figure}

As illustrated in Fig.~\ref{fig:decay}, to identify the critical point for a given $\tilde{g}$, we start with a highly coherent 2d gas 
and measure $n(k)$ after holding the cloud in the trap for a variable time $t$. 
During the hold time, the atom number $N$ slowly decays through various inelastic processes~\cite{Note3}, while the elastic-collision rate ($\approx 0.2\,N \, \tilde{g}^2~{\rm s}^{-1}$) remains sufficiently high to ensure that the gas is in quasi-static equilibrium. 
To measure $n(k)$, we employ the ``momentum focussing" technique~\cite{Shvarchuck:2002,Tung:2010,Murthy:2014,Ries:2014}. We turn off just the tight $z$ confinement, so the rapid vertical expansion (predominantly driven by the zero-point motion along $z$) removes all the interaction energy on a timescale $1/\omega_z \ll 1/\omega_r$. The subsequent horizontal ideal-gas evolution in the remaining in-plane harmonic potential reveals $n(k)$ as the spatial distribution after a quarter of the trap period. We probe this distribution by absorption imaging along $z$ [see Fig.~\ref{fig:decay}(a)].

Our $k$-space imaging resolution, $\Delta k \approx 0.4~\mu$m$^{-1}$, sets the largest distance over which we can probe coherence to $L=2\pi/\Delta k \approx15~\mu\text{m}$, which is much larger than $\lambda \approx 0.7~\mu$m.
To probe coherence on this lengthscale, we simply monitor the peak value of the momentum distribution, $P_0$, without making any theoretical assumptions about the exact shape of $n(k)$ at low $k$. To get the corresponding atom number $N$ we do a simple summation over the image. Importantly, we eliminate the systematic error due to the uncertainty in the absorption-imaging cross section by independently calibrating our imaging system through measurements of the BEC critical point in a 3d gas~\cite{Note4}.

In Fig.~\ref{fig:decay}(b) we show a typical evolution of $P_0$ and $N$ (here $\tilde{g}=0.28$).
While $N$ decays smoothly, $P_0$ shows two distinct regimes, which allows us to identify the critical hold time $t_c$ and the corresponding $N_c$.  
We note that even for $N$ significantly below $N_c$ the peak of $n(k)$ rises above a Gaussian fitted to the wings of the distribution, indicating some coherence on a lengthscale $> \lambda$~\cite{Clade:2009, Plisson:2011}. The smooth evolution of such non-Gaussian ``peakiness" of $n(k)$ does not reveal a phase transition~\cite{Plisson:2011}, and only $P_0$ corresponding to $L \gg \lambda$ shows a clear change in behaviour at a well-defined $N_c$~\cite{Note5}.
Our large $L$ is still small compared to the thermal diameter of the cloud, $2\sqrt{\kB T/\left(m \omega_r^2\right)} \approx 50~\mu$m, so the observed $N_c$ is closely linked to the occurrence of a phase transition in the centre of the trap~\cite{Note4}.

For comparisons with theory, we also fit $\mu$ and $T$ to each $n(k)$ image.
Unlike in 3d, in 2d interactions affect $n(k)$ appreciably even in the normal state, and near the critical point it is in general insufficient to treat them at a mean-field (MF) level. However, beyond-MF correlations primarily affect the highly-populated low-$k$ states~\cite{Holzmann:2010}. We restrict our fits to the high-$k$ wings of the distribution ($\hbar^2 k^2 > 2\tilde{g}m\kB T$), where we expect the beyond-MF effects to be small, and still carefully include the effects of interactions at a MF level~\cite{Note4}. Following \cite{Hadzibabic:2008}, we also account for the residual thermal occupation of the axial excited states and the interaction-induced deformation of the axial eigenstates.

\begin{figure}[tbp] 
\includegraphics[width=1\columnwidth]{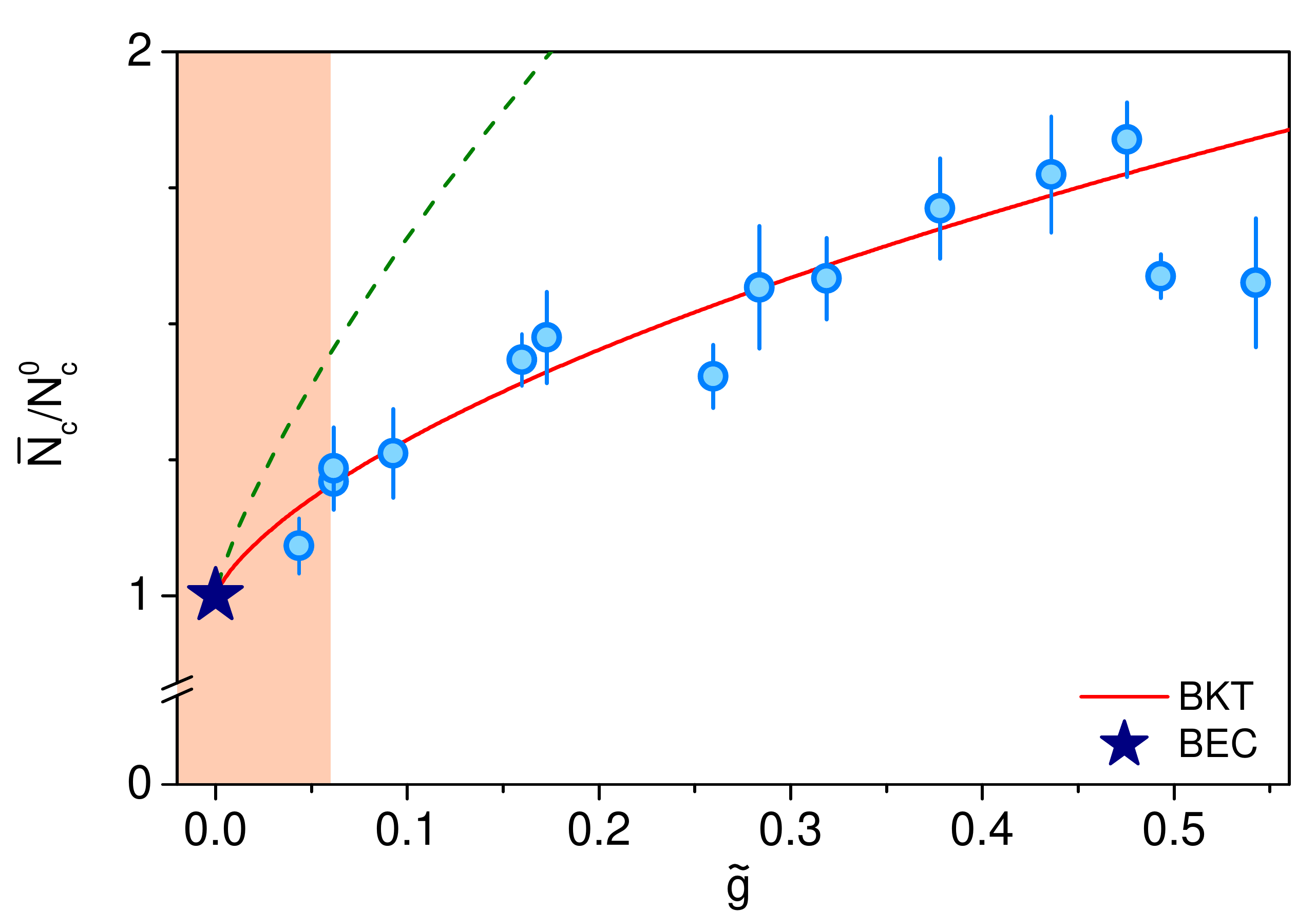} 
\caption{(color online) Critical atom number as a function of the interaction strength $\tilde{g}$.  All numbers are scaled to the ideal-gas BEC critical number $N_\text{c}^0$, defined in Eq.~(\ref{eq:Nc2D}). Solid line is the classical-field BKT prediction of Eq.~(\ref{eq:NcHolzmann}), without any free parameters. Dashed line is an approximation which neglects suppression of density fluctuations in the normal state. The star $\left(\star\right)$ denotes the critical point for BEC, which only occurs in the ideal-gas limit. The shaded region, $\tilde{g}<0.06$, indicates the regime in which our measurements stop being reliable (see text). Error bars are statistical.}
\label{fig:Nc}
\end{figure}

In Fig.~\ref{fig:Nc} we summarise our measurements of the critical atom number for a wide range of interaction strengths. To compare our data with the strictly-2d theoretical calculations, we correct the observed ``raw" $N_c$ by subtracting the calculated population of the excited axial states~\cite{Note4}. We scale this corrected critical number, $\bar{N}_c$, to the BEC critical atom number, $N_c^0$ of Eq.~(\ref{eq:Nc2D}), and plot it versus $\tilde{g}$.
Our $\Delta k$-limited value of $L$ imposes a lower bound on $\tilde{g}$ for which we can reliably identify the critical point. In the absence of any phase transition, in a weakly-interacting degenerate gas $g_1(r) \sim \exp(-r/\ell_0)$, with $\ell_0=\lambda \exp(D/2)/\sqrt{4\pi}$~\cite{Hadzibabic:2011}. We thus do not expect our measurements to reliably identify $N_c$ if $\ell_0 > L$ for some $D< D_\text{BKT}$. This occurs for $\tilde{g} < 380 \, \lambda^2 / (4\pi L^2) \approx 0.06$, indicated by the shaded area in Fig.~\ref{fig:Nc}. 
Our measurements also stop being reliable for $\tilde{g} \gtrsim 0.5$; in that regime our MF temperature fits are restricted to very high $k$ values, which are affected by the anharmonicity of the optical trap.
The error bars in Fig.~\ref{fig:Nc} are statistical, while the systematic uncertainty in $\bar{N}_\text{c}/N_\text{c}^0$ is $\lesssim 0.2$~\cite{Note4}. 

Without any free parameters, we find generally excellent agreement with the prediction of Ref.~\cite{Holzmann:2010}:
\begin{equation}
\frac{N_\text{c}^\text{BKT}}{N_\text{c}^0}\approx 1+\frac{3\tilde{g}}{\pi^3}\ln^2 \left(\frac{\tilde{g}}{16}\right) +\frac{6\tilde{g}}{16\pi^2}\left[15+\ln\left(\frac{\tilde{g}}{16}\right)\right] ,
\label{eq:NcHolzmann}
\end{equation} 
which is based on fixing the phase-space density in the trap centre to $D_\text{BKT}$ and integrating a uniform-system equation of state over the trap, using classical-field results of Ref.~\cite{Prokofev:2002}. 

The agreement with Eq.~(\ref{eq:NcHolzmann}) over a very broad range of interaction strengths, and the proximity of our lowest reliable $\tilde{g}$ values to zero, allow us to conclude that the critical atom number, without any free parameters, indeed smoothly converges onto the BEC result of Eq.~(\ref{eq:Nc2D}). 

It is instructive to also compare our data with the approximation $N_\text{c}^\text{BKT}/N_\text{c}^0 = 1+3\tilde{g}D_\text{BKT}^2/\pi^3$~~\cite{Holzmann:2007a,Holzmann:2008a}, shown by the dashed line in Fig.~\ref{fig:Nc}. Here, the critical phase-space density is again set to $D_\text{BKT}$, but the suppression of bosonic fluctuations in the normal state is neglected, i.e. the density profile is calculated using MF theory with an interaction potential $2gn(r)$, where $g=(\hbar^2/m)\tilde{g}$. Our data strongly exclude this result, confirming the importance of the suppression of density fluctuations near the critical point even for our lowest $\tilde{g}$ values.

For a more direct comparison with the uniform-system theory, we also consider the critical chemical potential for the onset of coherence. Like $N_c$ in Fig.~\ref{fig:decay}, $\mu_\text{c}$ is experimentally defined via the critical hold time $t_c$. The classical-field simulations~\cite{Prokofev:2001} predict $D_\text{BKT}$ to be reached for $\mu_\text{c}^\text{BKT}= \kB T \left(\tilde{g}/\pi\right)\ln\left(13.2/\tilde{g}\right)$, which reduces to the BEC prediction, $\mu_c=0$, for $\tilde{g} =0$.  

\begin{figure}[tbp] 
\includegraphics[width=1\columnwidth]{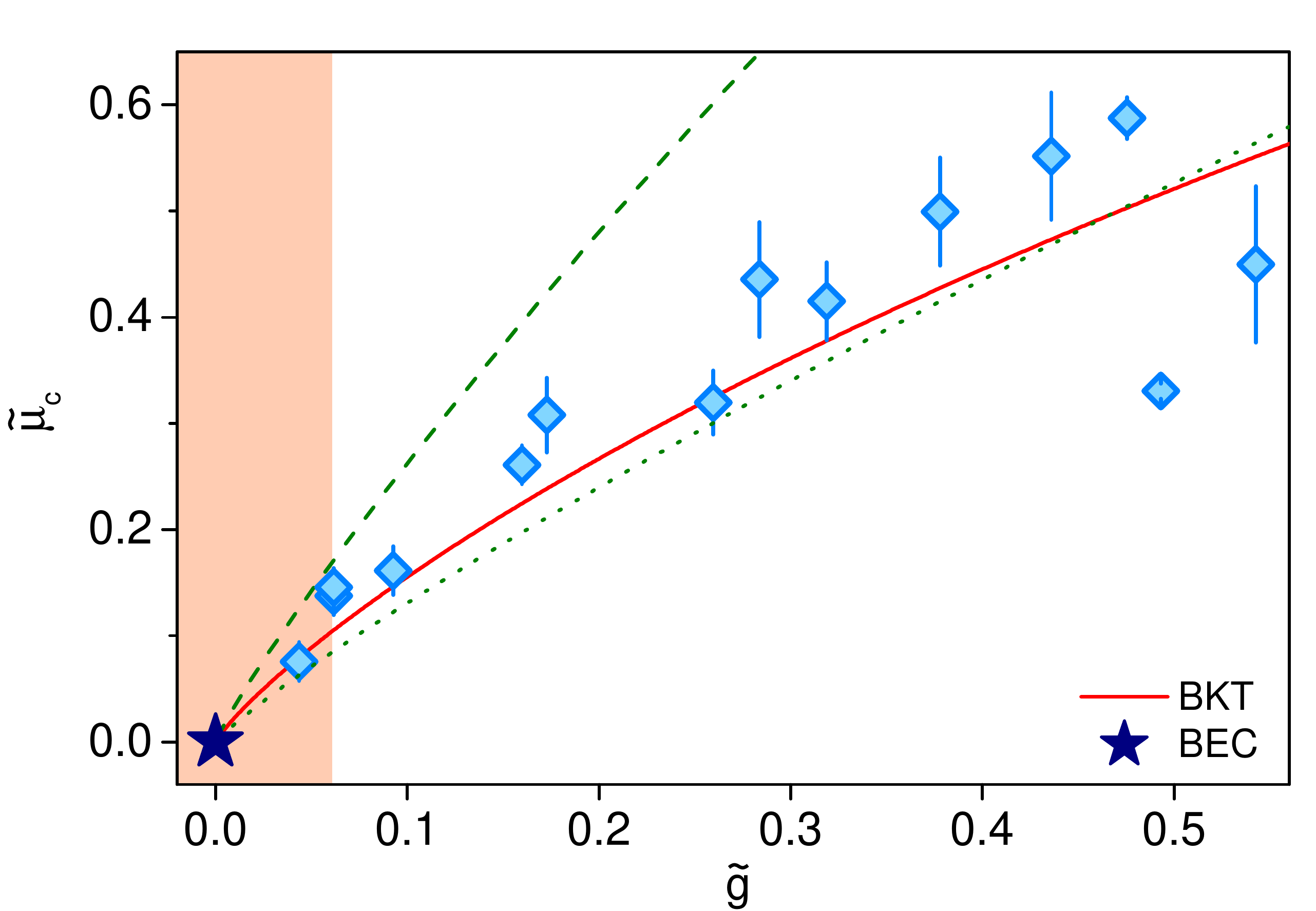}
\caption{(color online) Critical chemical potential as a function of the interaction strength $\tilde{g}$. 
Solid line is the classical-field BKT prediction. Dashed line is derived assuming a fully fluctuating Bose gas, while dotted line corresponds to a complete suppression of density fluctuations. Error bars are statistical.}
\label{fig:mu}
\end{figure}

In Fig.~\ref{fig:mu} we plot $\tilde{\mu}_\text{c} = \mu_\text{c}/(\kB T)$ versus $\tilde{g}$, and again observe generally good agreement with the classical-field prediction (solid line), all the way down to $\tilde{g}\approx 0.06$, i.e. very close to the expected BEC limit. The small systematic difference between the data and the theory is comparable to our systematic uncertainty in $\tilde{\mu}_\text{c}$, of $\sim 0.05$~\cite{Note4}.

We also compare our data with two intuitive approximations to $\tilde{\mu}_c$. We consider interaction potentials $\gamma gn$ with $\gamma=2$, corresponding to a fully fluctuating Bose gas, and $\gamma=1$, corresponding to a complete suppression of density fluctuations. In both of these extremes one can analytically write $D_\gamma\left(\tilde{\mu}\right)=-\ln\left[1-\exp\left(\tilde{\mu}-\gamma g n/(\kB T)\right)\right]$~\cite{Hadzibabic:2011}. Defining $\tilde{\mu}_\text{c}^\gamma$ so that $D_\gamma (\tilde{\mu}_\text{c}^\gamma)=D_\text{BKT}$ we obtain the dashed ($\gamma=2$) and dotted ($\gamma=1$) lines in Fig.~\ref{fig:mu}. 
Generally $\gamma=1$ provides a better approximation, 
highlighting how strong the suppression of density fluctuations in the normal state is. 

Finally, we note that in previous experiments~\cite{Hung:2011,Ha:2013}, on the in-trap equation of state, $\tilde{\mu}_c$ was deduced by defining it so as to satisfy the theoretical expectation~\cite{Prokofev:2001,Prokofev:2002} that the phase-space density should be a universal function of $(\tilde{\mu}-\tilde{\mu}_\text{c})/\tilde{g}$. 
Our measurements of $\tilde{\mu}_c$ defined through the emergence of extended coherence show that the two definitions indeed lead to very similar values.

In conclusion, by studying the critical point for the emergence of extended coherence in a harmonically trapped 2d gas with tuneable interactions, we have quantitatively confirmed the predictions of classical-field theory and observed the expected unification of BKT and BEC transitions in the limit of vanishing interactions. The in-plane harmonic potential enables this observation. However, to quantitatively study the exact functional form of the slowly decaying correlations in a BKT superfluid, in the future it would be very interesting to study coherence of a tuneable 2d gas in a uniform potential~\cite{Gaunt:2013, Corman:2014,Chomaz:2015}. Just below $T_c$ this should reveal an interaction-strength-independent algebraic decay of the first-order correlation function, corresponding to a universal jump in the superfluid density~\cite{Nelson:1977}.

We thank Naaman Tammuz and Alex Gaunt for experimental contributions in the early stages of the project, Shannon Whitlock for help with image processing, and Andre Wenz, Markus Holzmann and Jean Dalibard for useful discussions. This work was supported by AFOSR, ARO, DARPA OLE, and EPSRC [Grant No. EP/K003615/1]. N.N. acknowledges support from Trinity College, Cambridge, R.P.S. from the Royal Society, and K.G.H.V. from DAAD.





\begin{thebibliography}{10}
\bibitem{NoteCrossover}
We estimate the width of the crossover region by noting that to observe interaction physics the interaction energy must be larger than $\hbar \omega_r$.
\end{thebibliography}


\end{document}